\documentclass[runningheads]{llncs}
\usepackage{graphicx}

\usepackage{tikz}
\usepackage{comment}
\usepackage{amsmath,amssymb} 
\usepackage{color}

\usepackage[breaklinks,colorlinks]{hyperref}

\usepackage[accsupp]{axessibility}  % 

\usepackage{caption}
\usepackage{subcaption}
\usepackage{ulem,gensymb}

\begin{document}

\pagestyle{headings}
\mainmatter

\title{BTDNet: a Multi-Modal Approach for Brain Tumor Radiogenomic Classification}

\titlerunning{BTDNet: a Multi-Modal Approach for Brain Tumor Radiogenomic Classification from mpMRI Scans}

\author{Dimitrios Kollias, Karanjot Vendal, Priyanka Gadhavi and  Solomon Russom 
\authorrunning{D. Kollias}
\institute{Queen Mary University of London, UK \\
\email{d.kollias@qmul.ac.uk}}}

%******************
\maketitle

\abstract{Brain tumors pose significant health challenges worldwide, with glioblastoma being one of the most aggressive forms. Accurate determination of the O6-methylguanine-DNA methyltransferase (MGMT) promoter methylation status is crucial for personalized treatment strategies. However, traditional methods are labor-intensive and time-consuming. This paper proposes a novel multi-modal approach, BTDNet, leveraging multi-parametric MRI  scans, including FLAIR, T1w, T1wCE, and T2 3D volumes, to predict MGMT promoter methylation status. BTDNet addresses two main challenges: the variable volume lengths (i.e., each volume consists of a different number of slices) and the volume-level annotations (i.e., the whole 3D volume is annotated and not the independent slices that it consists of). BTDNet consists of four components: i) the data augmentation one (that performs geometric transformations, convex combinations of data pairs and test-time data augmentation); ii) the 3D analysis one (that performs global analysis through a CNN-RNN); iii) the routing one (that contains a mask layer that handles variable input feature lengths), and iv) the modality fusion one (that effectively enhances data representation, reduces ambiguities and mitigates data scarcity). The proposed method outperforms by large margins the state-of-the-art methods in the RSNA-ASNR-MICCAI BraTS 2021 Challenge, offering a promising avenue for enhancing brain tumor diagnosis and treatment.
}

\section{Introduction}

Brain tumors are a complex and heterogeneous group of neoplasms that pose a significant health challenge worldwide. Despite considerable progress in our understanding of their molecular and genetic underpinnings, the diagnosis, prognosis, and treatment of brain tumors remain formidable tasks. In recent years, the integration of radiological imaging data with genomic information has emerged as a promising avenue in the field of neuro-oncology. This intersection of radiology and genomics, often referred to as "radiogenomics," has the potential to revolutionize the approach to brain tumor characterization and classification.

The inherent complexity and diversity of brain tumors, coupled with the limitations of traditional histopathological methods, have spurred the search for non-invasive, complementary approaches to improve the accuracy of diagnosis and prognosis. Radiogenomic analysis leverages advanced imaging techniques such as magnetic resonance imaging (MRI), positron emission tomography (PET) and computed tomography (CT) to extract a wealth of quantitative and qualitative imaging features. These features are then correlated with the genetic and molecular characteristics of brain tumors, offering insights into their underlying biology and clinical behavior.

Glioblastoma, the most aggressive and prevalent primary brain tumor in adults, remains a therapeutic challenge despite advances in neuro-oncology. The genetic and epigenetic heterogeneity within glioblastoma underscores the urgent need for precise biomarkers to guide personalized treatment strategies. Among these, the methylation status of the O6-methylguanine-DNA methyltransferase (MGMT) promoter has emerged as a critical determinant of response to temozolomide, the standard chemotherapy for glioblastoma. The predictive power of MGMT promoter methylation status in glioblastoma treatment outcomes underscores the significance of accurate and timely determination. 
%
%
%
%Artificial Intelligence (AI) techniques, with their capacity to analyze complex genetic data, offer a promising avenue for enhancing the prediction of MGMT promoter methylation status.
%
Traditionally, the determination of MGMT promoter methylation status has relied on labor-intensive and time-consuming laboratory techniques, such as methylation-specific polymerase chain reaction (MSP) and pyrosequencing.

The emergence of AI-driven methods promises to revolutionize this process by analyzing high-dimensional genomic and epigenomic data.
In this work, we utilize multi-parametric MRI (mpMRI) scans, to address the prediction of the MGMT promoter methylation status at the pre-operative baseline MRI scans. The mpMRI scans consist of different types of 3D volumes (which consitute different modalities for the developed method): a) Fluid Attenuated Inversion Recovery (FLAIR); b) native (i.e., T1-weighted pre-contrast or in other words T1w);  c) post-contrast T1-weighted (T1wCE); and d) T2-weighted (T2) 3D volumes. 

The challenges with using such 3D volumes of data for prediction of the MGMT promoter methylation status are two-fold: i) the annotations are at volume-level rather than at slice-level, i.e., there exists one annotation for the whole volume; ii) the volumes have variable lengths, i.e., volumes are 3D signals consisting of series of slices, i.e., 2-D images, and each volume consists of a different number of images. In this work, we will address and overcome these challenges with our proposed method.

Traditional approaches handle 3D signals using 3D CNN architectures that give one prediction per signal \cite{baba_2021,roberts_2021}; however such architectures are very complex with a large number of parameters and require to have been pre-trained with other large 3D databases. In the healthcare domain and the medical imaging world, large 3D databases -annotated for the purpose of interest- are not easy to find and in most of the cases are not publicly available due to privacy issues, regulatory frameworks and policies. 
Other traditional approaches make a hypothesis and assign the volume-level label to each slice of the volume and then employ CNN-RNN networks to train with the annotated slices. However, the fact that the whole volume has one label does not mean that each slice in the volume exhibits the MGMT promoter methylation status; it could be the case that only some slices display the MGMT promoter methylation status.

In the case where the volumes have variable lengths, traditional approaches use some ad-hoc strategies; they select a fixed volume length and either remove slices when a larger length is met (thus losing information that could be important for the final decision), or duplicate slices when the volume contains a smaller number of slices (this duplication  affects negatively the final decision as the model gets biased towards the repeating data) \cite{phan_2021,soares_2021,tangirala_2021,kollias2023deepp}. What is more, this ad-hoc way of selecting the fixed volume length needs to be tuned empirically in every different database.

In this paper we propose BrainTumorDetectionNetwork (BTDNet), a multi-modal approach for prediction of MGMT promoter methylation status. We develop a multi-modal approach as leveraging data from multiple modalities helps the method to: i) have a richer data representation, since different modalities provide complementary information; ii) reduce redundancy and ambiguity in data interpretation; iii) mitigate data scarcity.
%\textcolor{red}{ + pes eite edw eite sto telos gia novel data augmentation kai loss function}

BTDNet takes as input an mpMRI scan, i.e., a FLAIR 3D volume, a T1w 3D volume, a T1wCE 3D volume, and a T2 3D volume. BTDNet consists of four components: i) the Data Augmentation; ii) the 3D Analysis; iii) the Routing; and iv) the Modality Fusion. At first, while training, each input volume is passed through the Data Augmentation component. That component applies at first geometric transformations to the input volume and then performs a novel augmentation technique to it. Let us note that, in inference, we apply Test-Time Data Augmentation to the input volume. The transformed input volume is then processed by the 3D Analysis and Routing components. 

The 3D Analysis component consists of a CNN network acting as a feature extractor and is applied to each slice of the volume. The features are then fed to an RNN that captures temporal information within the slices of the same volume. The RNN's features are then passed to the Routing component,  consisting of a Mask layer, which dynamically selects specific RNN outputs, followed by a dense layer (i.e., fully connected  layer equipped with batch normalization \cite{santurkar2018does} and GELU \cite{hendrycks2016gaussian}), whose output features are then fed to the Modality Fusion component. The Mask layer is utilized so as to handle variable input feature lengths (due to variable slices per volume) when training the network.

The 3D Analysis and Routing components are identical and share the same weights for each of the four input 3D volumes (modalities). This means that there exist 4 identical 3D Analysis and Routing components.
The Modality Fusion component is fed with the outputs of the Routing component of all four modalities; these outputs are first concatenated and then fed to a dense layer that maps them to the same feature space. Then the output layer follows that performs the final classification. Our method is trained using an extension of the Focal Loss \cite{lin2017focal} for multi-class classification.

The rest of the sections of this paper are as follows. Section 2 provides a comprehensive review of related works that address the problem of prediction of the MGMT promoter methylation status. Section 3 presents the proposed method, along with its novelties. It further describes the utilized dataset and the pre-processing techniques that we used; finally it presents the  performance metrics used to evaluate our approach, as well as implementation details regarding our approach. Section 4 presents the rich experimental study in which we compare out method's performance to that of the state-of-the-art, as well as various ablation studies are presented. Finally, Section 5 provides a conclusion to our work.

\section{Related Work}
%A thorough analysis of several learning-based strategies has been conducted in the search for a precise method for predicting the MGMT methylation status from brain MRI data. In general, this task is viewed as a binary classification task that separates methylated from non-methylated states. In this field, well-known feature-based algorithms like SVM, RF, KNN, J48, NB, and XGBoost have been widely used \cite{le2020xgboost, korfiatis2016mri,kanas2017learning}. The advancement of modern deep learning paradigms, particularly the CNN and RNN models, has enhanced classifier evolution \cite{korfiatis2017residual, li2018multiregional, han2018mri, chen2020automatic, yogananda2021mri}.
In this section we describe the methods that participated in the RSNA-ASNR-MICCAI BraTS 2021 Challenge \cite{baid2021rsna} for classifying the tumor’s MGMT promoter methylation status from  pre-operative baseline mpMRI data of
2,040 patients. Let us note that all methods: i) tackled the variable volume lengths by either sub-sampling or duplicating slices; ii) tackled the annotation being at volume-level either by 3D CNN networks or by CNN-RNN in which the annotation was propagated to each slice within the volume; iii) developed either multimodal approaches (utilizing all modalities of the mpMRI scan) or unimodal ones (utilizing just the FLAIR modality).

The winning method of the RSNA-ASNR-MICCAI BraTS 2021 Challenge (denoted thereafter as 3D-Resnet10-Trick) employed a 3D CNN model utilizing the Resnet10 \cite{he2016deep} architecture with the FLAIR modality from the mpMRI \cite{baba_2021}. The model processed slices of size $256 \times 256$. A  technique named "Best Central Image Trick", was introduced to construct the 3D input volumes to the model. At first the 'best' slice was selected as the central image of the newly constructed 3D volume (the word 'best' denotes the image that contains the largest brain cutaway view). Then the 20 slices that exist before the 'best' slice and the 20 slices that exist after the 'best' slice are selected (the words 'before' and 'after' refer to the slices within the original FLAIR 3D volume). Finally all these slices are depth concatenated to form the input volume to the 3D-Resnet10.

The runner-up solution of the competition (denoted thereafter as EfficientNet-LSTM-mpMRI) leveraged a CNN-RNN achitecture in which   EfficientNet-B0 \cite{tan2019efficientnet} was the selected CNN model and LSTM \cite{hochreiter1997long} was the selected RNN model \cite{phan_2021}. EfficientNet-B0 was pre-trained and used as a feature extractor, whereas the LSTM was trainable. This approach was multimodal and utilized all four  modalities: FLAIR, T1w, T1wCE, and T2. A fixed temporal subsampling was performed to produce ten slices from each MRI modality. These slices were depth concatenated into a 4-channel map, with absent MRI types replaced by zero-filled channels. A 2D convolution converted the 4-channel image into a 3-channel feature map, aligning with EfficientNet's requirement.  Stratified K-fold cross-validation with \(K=5\) was employed, and ensemble predictions were derived by averaging outputs from all models.

The third-place solution (denoted thereafter as EfficientNet-Aggr) was a multimodal one comprised of four instances of  EfficientNet-B3 model, each dedicated to a specific modality \cite{soares_2021}. This approach utilized a stratified split based on patient IDs and classes within the training dataset to derive a validation set. 
Aggregation of results was performed on a per-patient basis, emphasizing the divergence between maximum predictions and the mean, and subsequently comparing the average and minimum predictions. The prediction associated with the most pronounced difference was retained. %This methodological choice underscored the importance of capturing the most distinguishing features in mpMRI scan types for accurate classification.

The fourth-placed solution (denoted thereafter as YOLO-EfficientNet)
combined Object Detection using YOLOv5 \cite{yolov5} and Classification via 2D and 3D EfficientNet variants \cite{roberts_2021}. Focusing on the T1wCE type, they employed a systematic seven-slice selection after resampling. The YOLOv5 object detection model, trained on hand-annotated images, identified tumor slices. For 2D classification, the best results were from the T1wCE axials with the EfficientNet B3 model, using diverse augmentations (centre cropping and adding noise to 3D data) and techniques like Test Time Augmentation (TTA) and power ensembling. In the absence of detectable tumors, all slices are globed together and passed on as 3D input. Model was trained with stratified 5-fold.

The fifth-place solution (denoted thereafter as stats-EfficientNet) \cite{tangirala_2021} at first sampled ten images from each modality. The mean of each type was computed to yield four 2D images. These images were subsequently depth concatenated into a 4-channel image, which was then processed through a 1x1 convolution bottleneck, resulting in a 3-channel feature map. This feature map was introduced to EfficientNet, to predict the MGMT value. Given the potential noise in the dataset due to its limited sample size, Taylor Cross Entropy loss was employed.

\section{Materials and Method}

In this section, at first, we present our proposed method, BTDNet, detailing all the components that it consists of (Data Augmentation;  3D Analysis; Routing; Modality Fusion) as well as the objective function that was used in its development. Then, we provide details on the utilized in this work dataset, which was used in the RSNA-ASNR-MICCAI BraTS 2021 Challenge. We also present and explain all data pre-processing steps that we followed. Next, we present the performance metric that we utilized to assess the performance of our method. Finally, we present all training implementation details corresponding to the development training and evaluation of our method.

\subsection{Proposed Method}

At first all input 3D volumes for each modality (FLAIR, T1w, T1wCE and T2) are fed to the Data Augmentation component. While training, this component applies a series of geometric transformations to the inputs and then performs a novel augmentation technique (MixAugment) to them. While testing, this component applies Test-Time Data Augmentation to the inputs.

Then all input 3D transformed volumes for each modality (FLAIR, T1w, T1wCE and T2) are padded with black slices to have lengths $t_1, t_2, t_3, t_4$ (e.g., each input FLAIR volume consists of $t_1$ slices, whereas each input T1w volume consists of $t_2$ slices). Each input 3D volume is first fed to the 3D Analysis component. This component consists of a CNN which performs local (per 2-D image/slice) analysis, extracting features  from each slice. Then these features are fed to an RNN, which is placed on top of the CNN, so as to capture their temporal dependencies. The CNN and RNN networks perform global (per 3D volume) analysis. The RNN output features are then fed to the Routing component, in which they are concatenated and  fed to a Mask layer. This step is essential since we have annotations at volume-level (not at slice-level) and thus we know that all slices (and not just independent slices) may convey important information for the final prediction of the network. This step is also important as it is the Mask layer that dynamically selects RNN outputs taking into account the input length, i.e., the 'true' number of slices of the currently analyzed volume. The output of the Mask layer is then fed to a dense layer which is equipped with batch normalisation and GELU.

Let us note that the above described procedure is followed for each modality. For example, the input 3D FLAIR volume is fed to the 3D Analysis component; the input 3D T1 volume is fed to the 3D Analysis component that is identical and shares the same weights with the corresponding component that the FLAIR modality was fed to. Then the output of the 3D Analysis component of the FLAIR modality is fed to the Routing component; also the output of the 3D Analysis component of the T1 modality is fed to the Routing component
which is identical and shares the same weights with the corresponding component of the FLAIR modality.

\begin{figure}[!h]
\centering
  \includegraphics[width=1\linewidth]{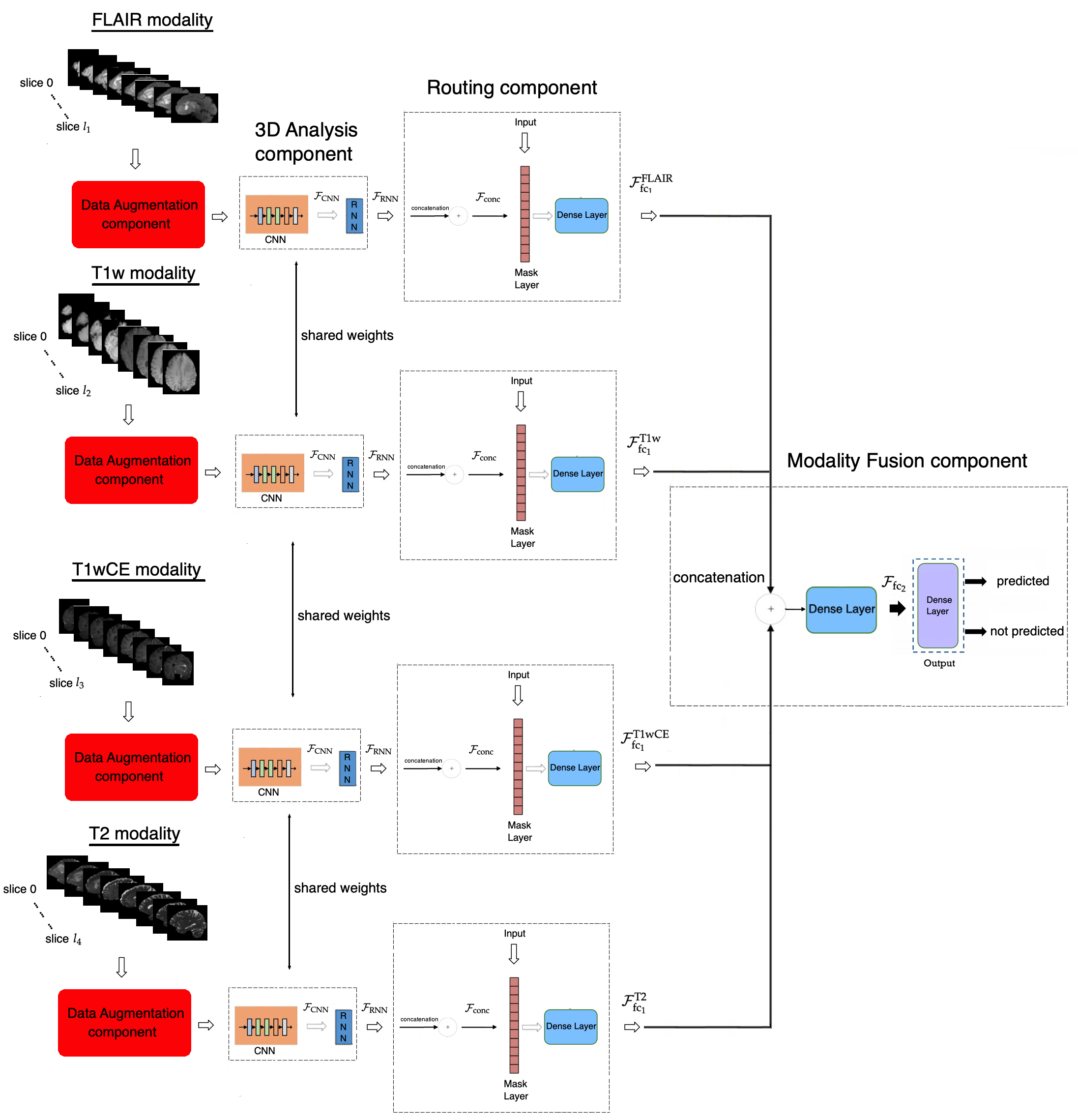}
  \caption{The proposed multimodal BTDNet that takes as input mpMRI 3D volumes (FLAIR, T1w, T1wCE and T2 modalities). BTDNet consists of the: i) Data Augmentation component (geometric transformations, adapted MixAugment, Test-Time Data Augmentation); ii) 3D Analysis component (CNN plus RNN); iii) Routing component (Mask layer, routing, dense layer); and iv) Modality Fusion component (dense layer, output layer).  }
  \label{methodology}
\end{figure}

The output features of the Routing component for each modality are fed to the Modality Fusion component. The  output features are concatenated and fed to a dense layer that maps them to the same feature space. 
%
%
%
%it is the Mask layer that handles the variable input video lengths with its routing mechanism that selects only the 'true' videoframes and not the duplicated ones. + that dynamically selects  RNN outputs taking into account the input length, i.e., the number of frames of the currently analyzed video.
%
%
Finally, the output layer follows providing the final classification.  In the following, we further explain in more detail each component of our proposed method.
Fig. \ref{methodology} gives an overview of our proposed framework, BTDNet.

\subsubsection{\underline{Data Augmentation Component}}
\vspace{0.1cm}

During training, when batches of 3D volumes ($X$) are sampled (i.e., all slices within the 3D volumes are sampled), at first we perform geometric transformations such as random horizontal flip and rotation and create transformed batches of 3D volumes ($T_X$). Let us note that we do not perform the same geometric transformation to all slices within the same volume; to the contrary, we apply a different geometric transformation to each slice of the same volume (e.g. one slice is augmented with horizontal flip and 10\degree rotation, whereas its following slice is augmented only with -4\degree rotation). 
After that, we extend MixAugment \cite{psaroudakis2022mixaugment} to be applied to these transformed batches of 3D volumes. MixAugment is a simple and data-agnostic data augmentation routine that trains a method on convex combinations of pairs of examples and their labels. It extends the training distribution by incorporating the prior knowledge that linear interpolations of feature vectors should lead to linear interpolations of the associated targets. 
We extend MixAugment to construct virtual training examples/3D volumes $(\tilde{X},\tilde{y})$ (one for each modality) and the method is trained concurrently on both real (r) and virtual (v) examples/3D volumes. The virtual examples for e.g. the FLAIR modality are constructed as follows (similarly are constructed the virtual examples for the other modalities):
\begin{align}
\tilde{X} &= \lambda T_{X_i}+(1-\lambda) T_{X_j}  \nonumber \\ 
\tilde{y} &= \lambda y_i+(1-\lambda)y_j
\label{eq:mixup}
\end{align}

\noindent where: $T_{X_i}\in \mathcal{R}^{H \times W \times t_1}$ and $T_{X_j}\in \mathcal{R}^{H \times W \times t_1}$ are two random FLAIR modality inputs (i.e., 3D volumes) with H and W being the height and width of each slice (i.e., 2D image) and $t_1$ being the total number of slices -after padding- within the volume; 
$y_i$ and $y_j$ $\in \{0,1\}^2$ are their corresponding one-hot labels;  
$\lambda \thicksim \mathrm{B}(\alpha,\alpha) \in [0,1]$ (i.e., Beta distribution) for $\alpha \in (0, \infty)$. 

%\textcolor{red}{For simplicity of presentation, the figure and the previous component sections do not contain details and mention about the data augmentation. However, data augmentation has been performed as we described before.}

During inference, we apply Test-Time Data Augmentation. For a given input 3D volume ($X_i$), we create three transformed versions of it: i) one version is when only horizontal flip is performed ($F_{X_i}$); ii) another version is when random rotation is performed ($R_{X_i}$); and iii) another version is when both horizontal flip and random rotation is performed ($FR_{X_i}$). We feed these three versions along with the non-augmented 3D volume to the method which gives a prediction for each version ($p^{F_{X_i}}, p^{R_{X_i}}, p^{FR_{X_i}}, p^{X_i}$, respectively). The final prediction ($p_{\text{final}}$) is the sum of these four outputs ($p_{\text{final}} = p^{F_{X_i}} + p^{R_{X_i}} + p^{FR_{X_i}} + p^{X_i}$). Let us note that we use the logits rather than the probabilities.

\subsubsection{\underline{3D Analysis Component}}
\vspace{0.1cm}

After all input 3D volumes for each modality (FLAIR, T1w, T1wCE and T2) have been transformed and then padded  to have lengths $t_1, t_2, t_3, t_4$, a CNN is applied to each of their slices, performing local (per 2-D image/slice) analysis, extracting features from each slice; e.g. for a slice of a 3D input volume from the FLAIR modality, the extracted features from the CNN are: $\mathcal{F_{\text{CNN}}} \in \mathcal{R}^{H \times W \times D}$, with $W$, $H$, $D$ being the feature map's width, height and depth, respectively. Then these features are fed to an RNN, which is placed on top of the CNN so as to capture temporal information and dependencies between consecutive slices within the same volume (for each modality). The RNN analyzes the features of the whole volume sequence, sequentially moving from slice $0$ to slice $t_1$, if the input volume is for the FLAIR modality (and similarly for the other modalities). 
As shown in Fig. \ref{methodology}, we get RNN features corresponding to each slice, from $0$ to $t_1$;
e.g. for a slice of a 3D input volume from the FLAIR modality, the extracted features from the RNN are: $\mathcal{F_{\text{RNN}}} \in \mathcal{R}^{V}$, with $V$ being the feature vector that is outputted by the RNN; thus for the whole 3D input volume from the FLAIR modality the extracted features from the RNN are: $\mathcal{F_{\text{RNN}}} \in \mathcal{R}^{V \times t_1}$.
These RNN features are then fed to the Routing component, described next. In total, the CNN plus RNN perform global (per 3D volume) analysis.

\subsubsection{\underline{Routing Component}}
\vspace{0.1cm}

In this component, at first, the RNN features corresponding to the whole input 3D volume are concatenated since our target is the prediction of the MGMT promoter methylation status using the whole volume, similarly to the annotations provided in the utilized dataset. The concatenated features ($\mathcal{F_{\text{conc}}} \in \mathcal{R}^{V \cdot t_1}$) are then fed to the Mask layer. The original (before padding) length $l_1$ of the input FLAIR volume (similarly $l_2, l_3, l_4$ for the T1w, T1wCE, T2 volumes) is transferred from the input to the Mask layer to inform the routing process. During model training, the routing mechanism performs dynamic selection of the RNN outputs, selecting as many of them as denoted by the length $l$ of the input volume, to keep their values, while zeroing the values of the rest RNN outputs and thus routing only the selected ones into the following dense layer. 

This dense layer is equipped with batch normalization and GELU. We use GELU due to its advantages over ReLU and its variants.
The dense layer learns to extract high level information from the concatenated RNN outputs. During training, we update only the weights that connect the dense layer neurons with the RNN outputs routed in the concatenated vector by the Mask layer. The remaining weights are updated whenever (i.e., in another input volume) respective RNN outputs are selected in the concatenated vector by the Mask layer. Objective function minimization is performed, as in networks with dynamic routing, by keeping the weights that do not participate in the routing process constant, and ignoring links that correspond to non-routed RNN outputs. Finally, the output of the dense layer ($\mathcal{F}_{\text{fc}_1} \in \mathcal{R}^{V'}$) is fed to the Modality Fusion component as explained next.

\subsubsection{\underline{Modality Fusion Component}}
\vspace{0.1cm}

At first, the output features of the dense layer for the FLAIR modality $\mathcal{F}_{\text{fc}_1}^{\text{FLAIR}} \in \mathcal{R}^{V'}$ are concatenated with the corresponding output features of the dense layer for the other modalities, T1w, T1wCE and T2 ($\mathcal{F}_{\text{fc}_1}^{\text{T1w}} \in \mathcal{R}^{V'}$, $\mathcal{F}_{\text{fc}_1}^{\text{T1wCE}} \in \mathcal{R}^{V'}$, $\mathcal{F}_{\text{fc}_1}^{\text{T2}} \in \mathcal{R}^{V'}$), respectively. These features have already been normalised in the Routing component where batch normalization was performed on the dense layers' outputs for each modality; therefore  features coming from different modalities are all within the same range. The concatenated features are then  fed to another dense layer ($\mathcal{F}_{\text{fc}_2} \in \mathcal{R}^{V''}$) -equipped with GELU- that maps them to the same feature space. This step is important as  leveraging and fusing information from multiple modalities helps the method to have a richer data representation, to reduce redundancy and ambiguity in data interpretation and to mitigate data scarcity. 
Finally the output layer follows which consists of two units and gives the final classification for the MGMT promoter methylation status.

\subsubsection{\underline{Objective Function}}
\vspace{0.1cm}

For the objective function we built upon the Focal Loss (FL) for multi-class classification, which is defined as follows:
\begin{align}
&\mathcal{L}_{FL} =  \sum_{i=1}^{batch} \bigg [ - \alpha (1-p)^\gamma \text{ log}p - (1-\alpha)p^\gamma \text{ log}(1-p)        \bigg ]
%
%
%\mathcal{L}_{CCE}^{v} + \mathcal{L}_{CCE}^{r_i} + \mathcal{L}_{CCE}^{r_j} 
\label{eq:focal}
\end{align}

\noindent where:
$p$ is the method’s estimated probability for the positive class;
$\alpha$ balances the importance of positive/negative examples;
$(1-p)^\gamma$ and $p^\gamma$ are modulating factors that reduce the loss contribution from easy examples and extends the range in which an example receives low loss; these modulating factors consist of the tunable focusing parameter $\gamma \ge 0$ that smoothly adjusts the rate at which easy examples are down-weighted.

As explained in the MixAugment strategy above, in each training iteration, the method is fed with both $T_{X_i}$ and $T_{X_j}$, and the constructed virtual volume $\tilde{X}$ (of Eq. \ref{eq:mixup}). Therefore the overall objective function consists of the sum of the focal losses for the real (r) and virtual (v) volumes:
\begin{align}
&\mathcal{L}_{total} =  \mathcal{L}_{FL}^{v} + \mathcal{L}_{FL}^{r_i} + \mathcal{L}_{FL}^{r_j}
\label{eq:total}
\end{align}

\subsection{Dataset}

\begin{figure}[!h]
    \centering
    \includegraphics[width=1.3cm]{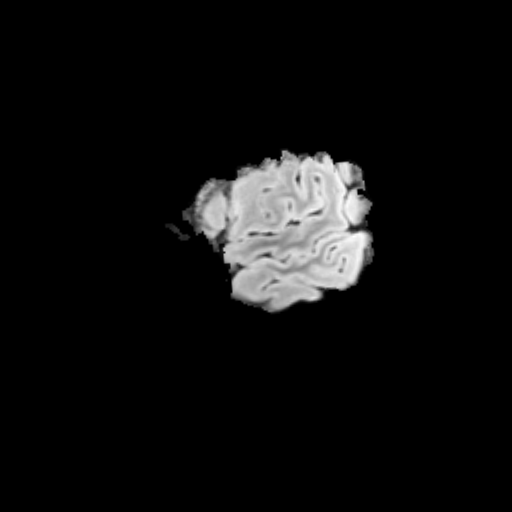}
    \includegraphics[width=1.3cm]{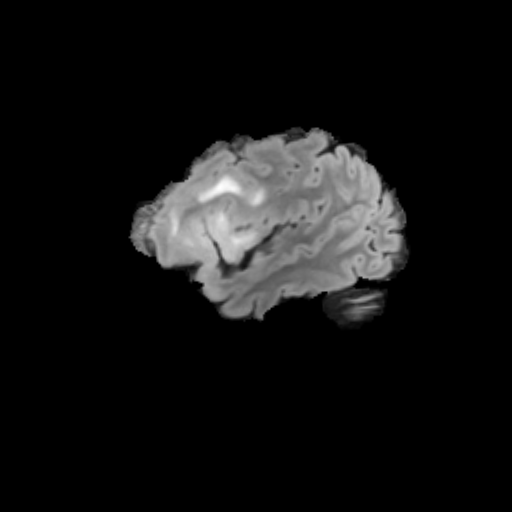}
    \includegraphics[width=1.3cm]{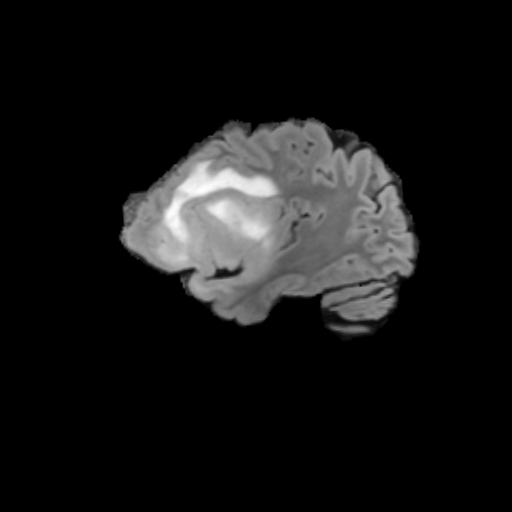}
    \includegraphics[width=1.3cm]{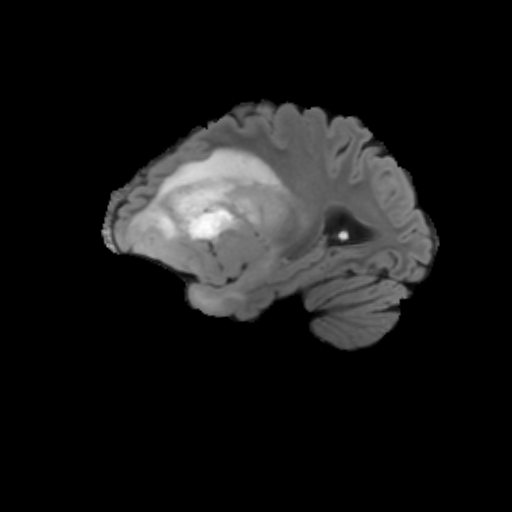}
    \includegraphics[width=1.3cm]{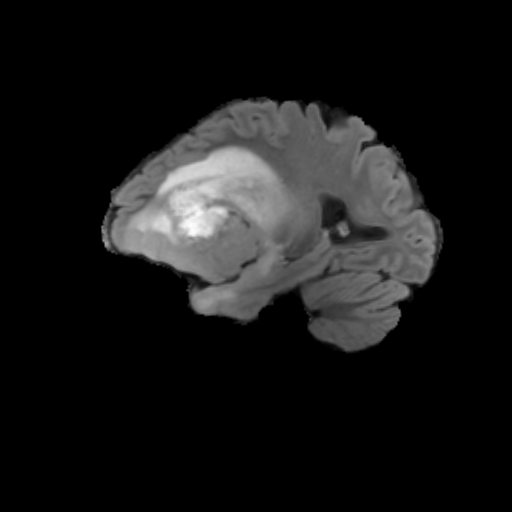}
    \includegraphics[width=1.3cm]{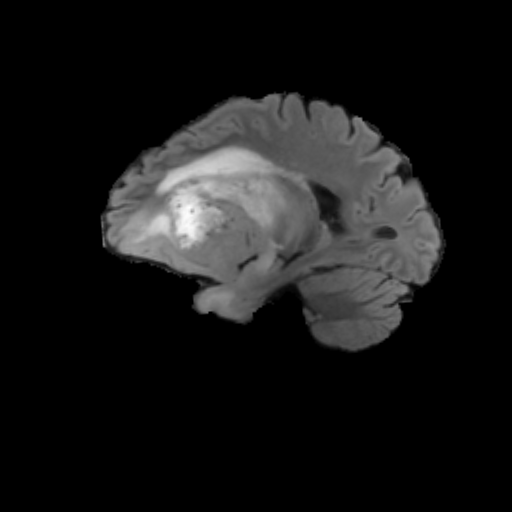}
    \includegraphics[width=1.3cm]{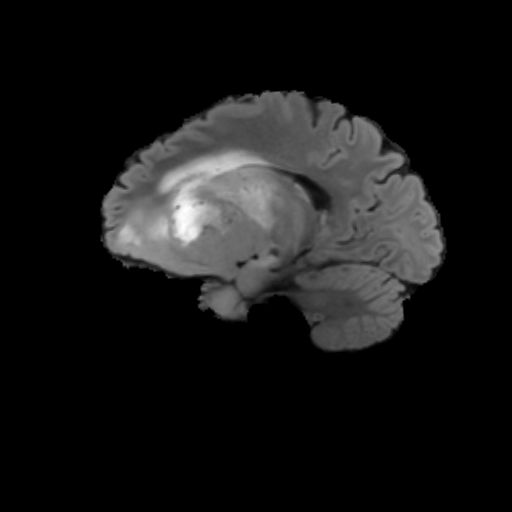}
    \includegraphics[width=1.3cm]{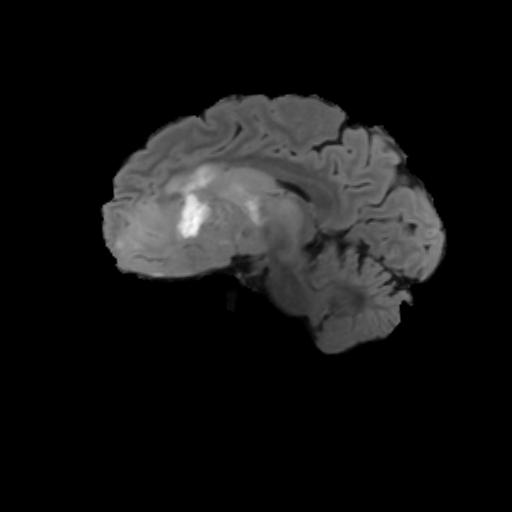} 
    \\
    \includegraphics[width=1.3cm]{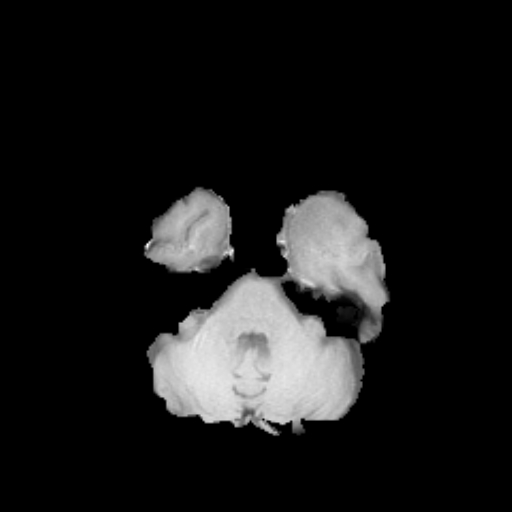}
    \includegraphics[width=1.3cm]{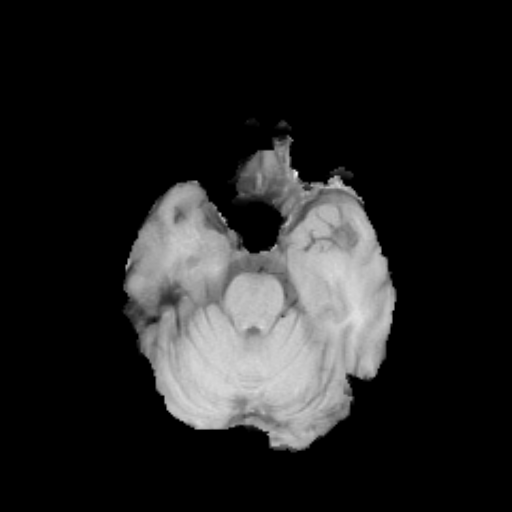}
    \includegraphics[width=1.3cm]{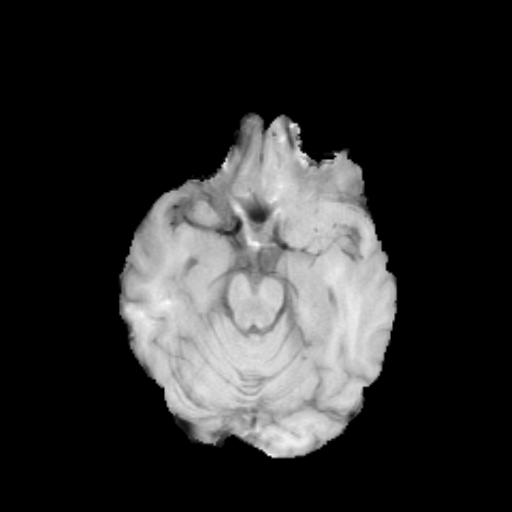}
    \includegraphics[width=1.3cm]{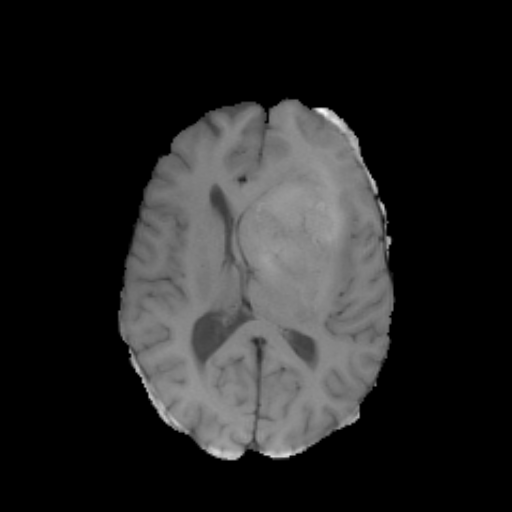}
    \includegraphics[width=1.3cm]{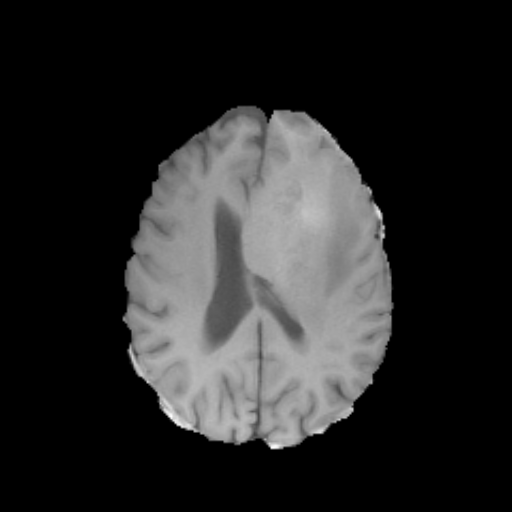}
    \includegraphics[width=1.3cm]{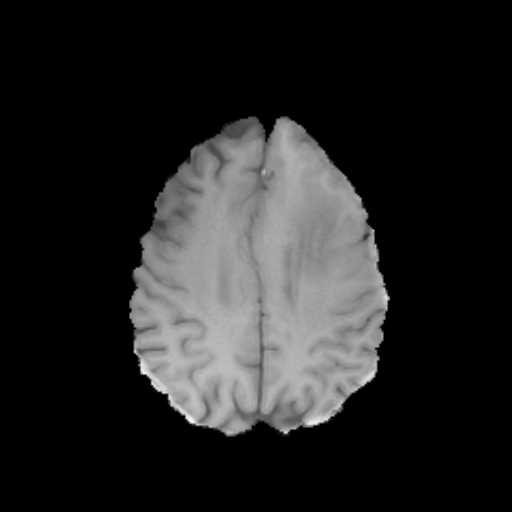} 
    \includegraphics[width=1.3cm]{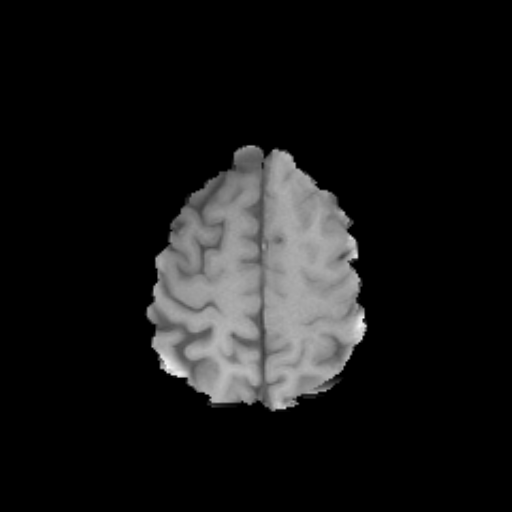} 
    \includegraphics[width=1.3cm]{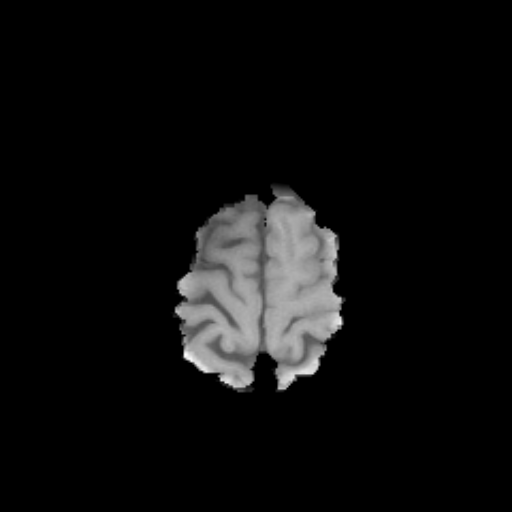} 
    \\
    \includegraphics[width=1.3cm]{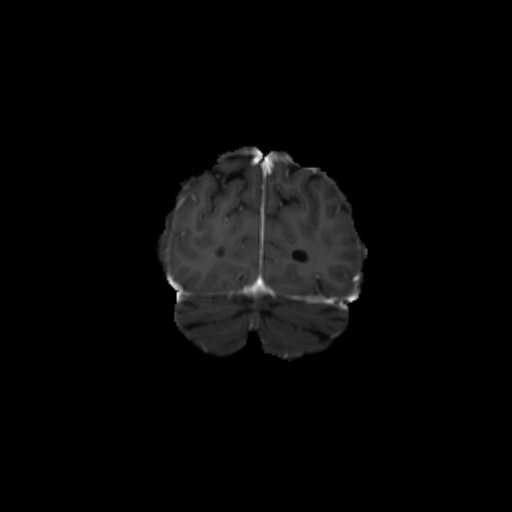}
    \includegraphics[width=1.3cm]{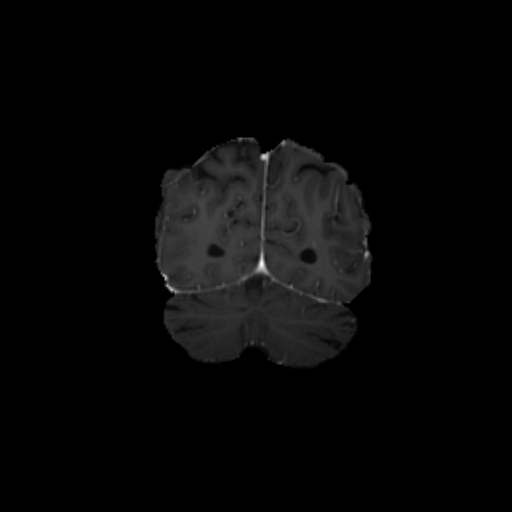}
    \includegraphics[width=1.3cm]{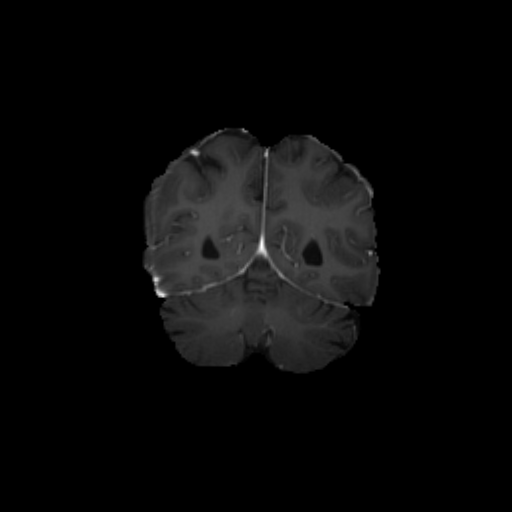}
    \includegraphics[width=1.3cm]{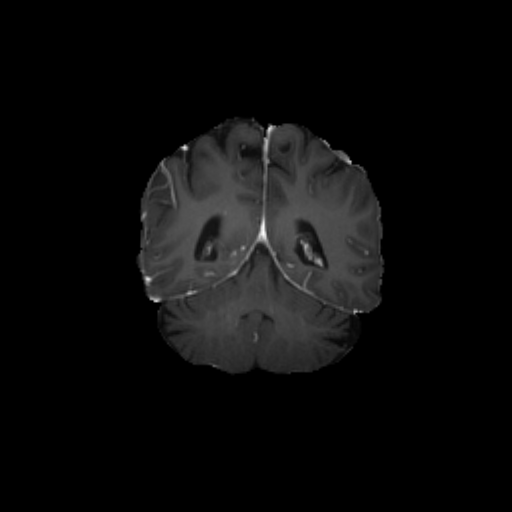}
    \includegraphics[width=1.3cm]{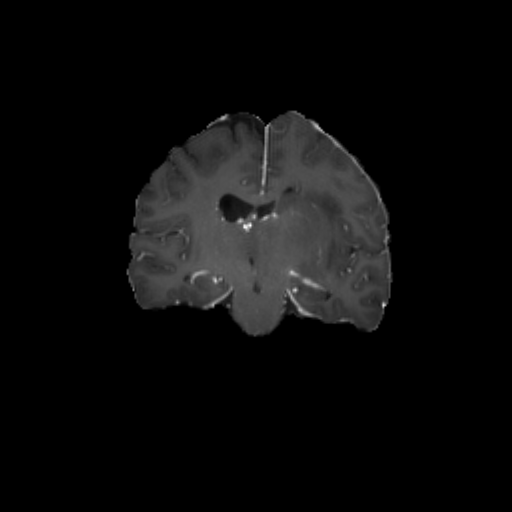}
    \includegraphics[width=1.3cm]{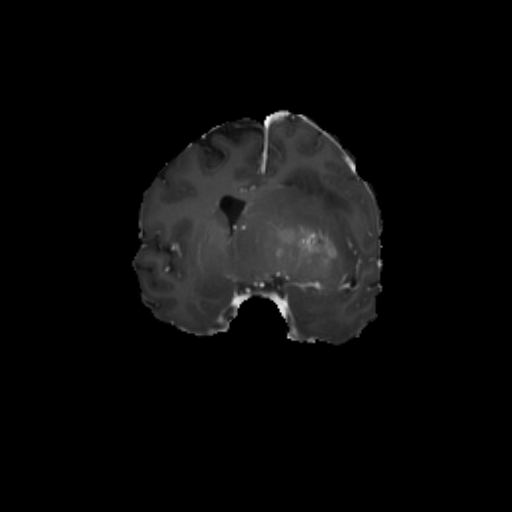}
    \includegraphics[width=1.3cm]{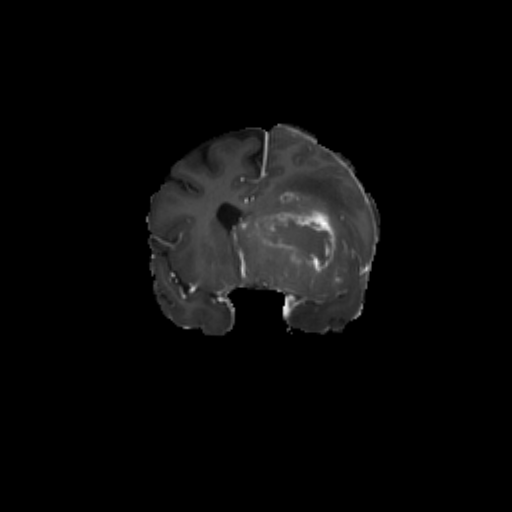} 
    \includegraphics[width=1.3cm]{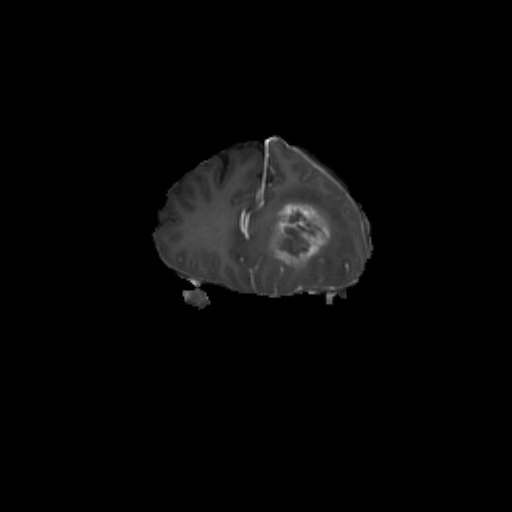} 
    \\
    \includegraphics[width=1.3cm]{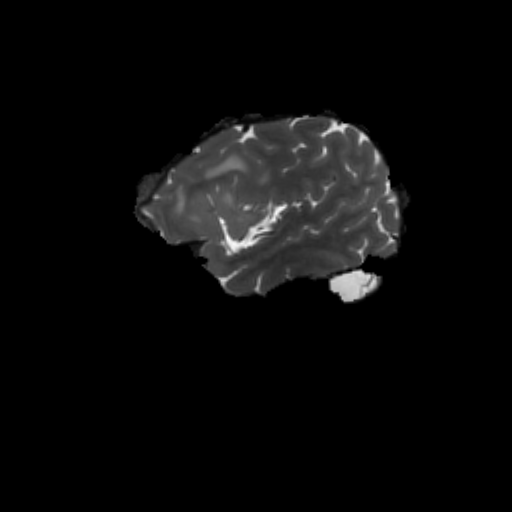}
    \includegraphics[width=1.3cm]{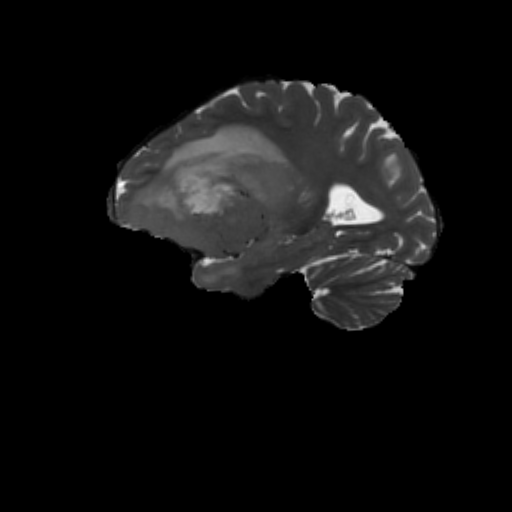}
    \includegraphics[width=1.3cm]{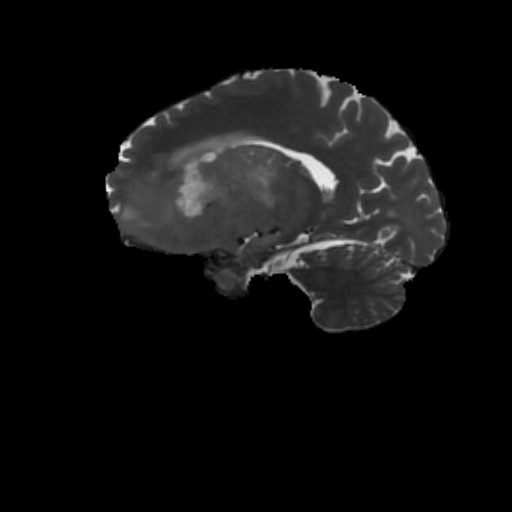}
    \includegraphics[width=1.3cm]{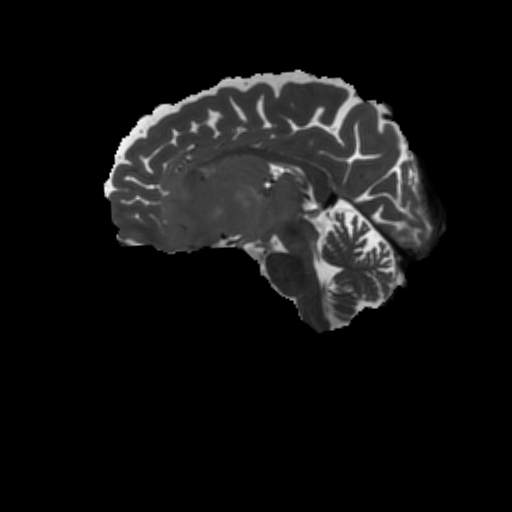}
    \includegraphics[width=1.3cm]{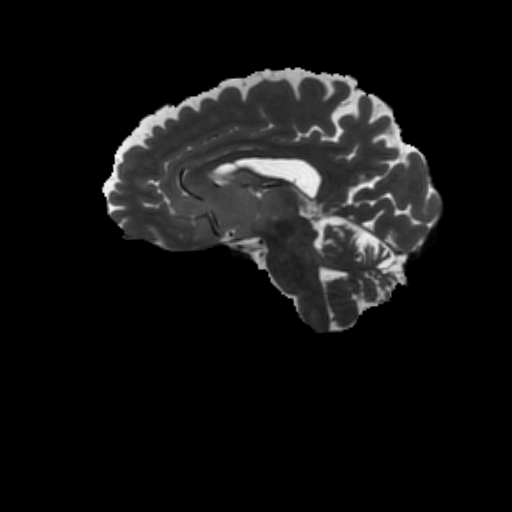} 
    \includegraphics[width=1.3cm]{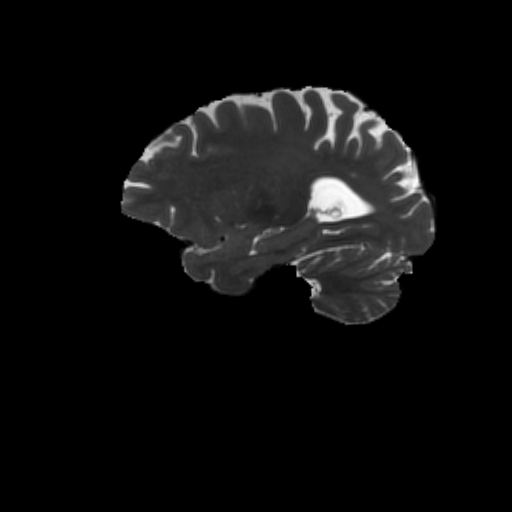} 
    \includegraphics[width=1.3cm]{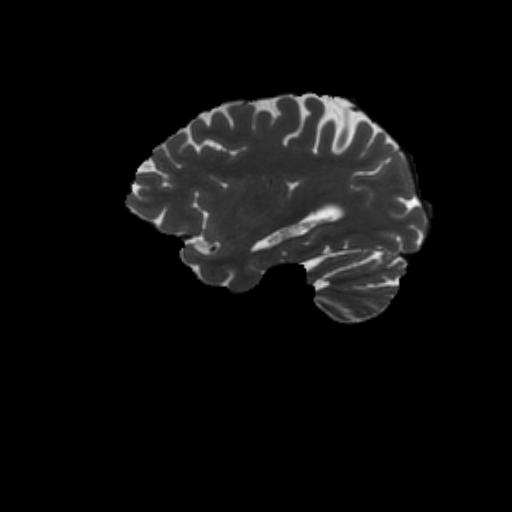} 
    \includegraphics[width=1.3cm]{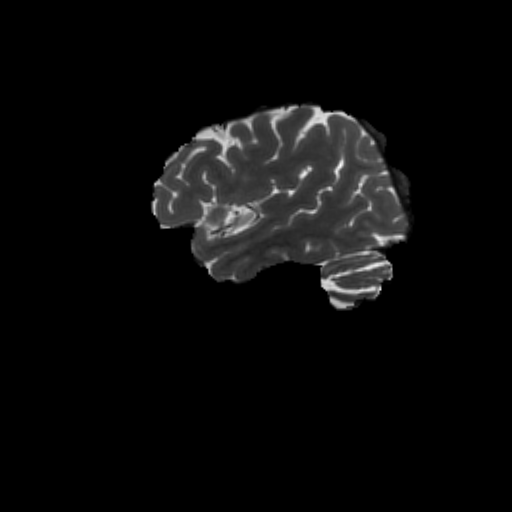} 

    \caption{Example of a whole mpMRI scan, where one can see some slices from: FLAIR modality (top row); T1w modality (second row);  T1wCE modality (third row); T2 modality (bottom row).}
    \label{datascan}
\end{figure}

\begin{figure}[!h]
    \centering
    \captionsetup{justification=centering}
    \includegraphics[width=7.3cm]{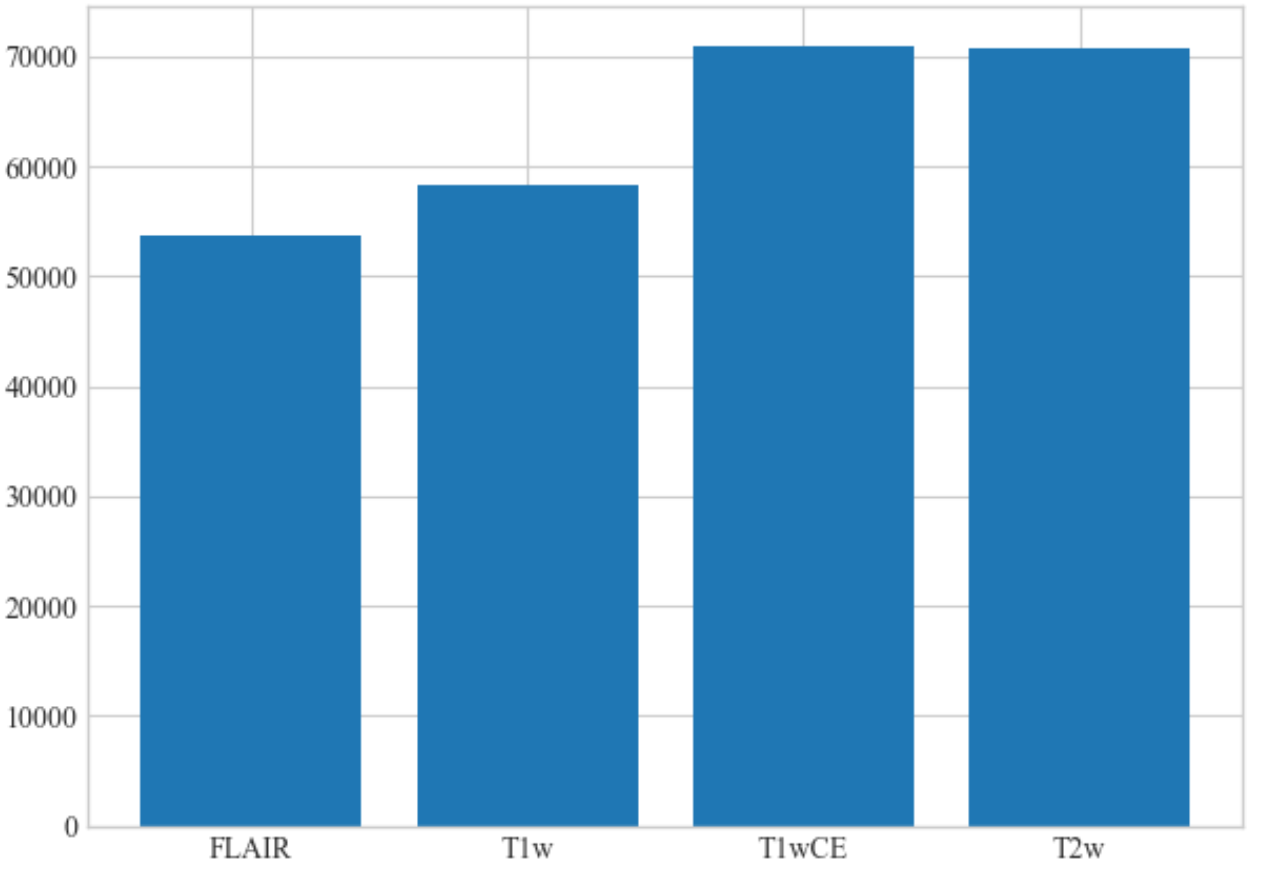}
    \caption{The total number of slices of each modality for all mpMRI scans.}
    \label{slices}
\end{figure}

The RSNA-MICCAI consortium has unveiled a dataset \cite{baid2021rsna} that comprises of multi-parametric Magnetic Resonance Imaging (mpMRI) scans (i.e. 3D volumes) from various institutions, annotated for the prediction of a specific genetic characteristic of glioblastoma, namely the MGMT promoter methylation status.
Each mpMRI scan consists of four modalities: Fluid attenuated inversion recovery (FLAIR), T1-weighted pre-contrast (T1w), T1-weighted post-contrast (T1wCE), and T2-weighted (T2). Each of these modalities offers a specific imaging perspective, e.g. the FLAIR modality provides imagery post cerebrospinal fluid suppression, where fluidic signals, such as water, are muted to emphasize other components. Fig. \ref{datascan} shows some slices from each modality of the same mpMRI scan. Each modality contains a varying number of slices. Fig. \ref{slices} illustrates the total number of slices of each modality for all mpMRI scans.
In total, the dataset comprises of 585 labeled samples, each with dimensions $512 \times 512$.  %It contains 3070 positive MRI scans (accounting for 57.5\%) and 2780 negative scans (constituting 52.5\%). 

%An important transformation of the RSNA-MICCAI dataset was executed by Jonathan Besomi \cite{besomi_2021}, where the data was converted from its original DICOM format to PNG. 

\subsection{Pre-Processing}

\begin{figure}[!h]
    \centering
    \captionsetup{justification=centering}
    \includegraphics[width=8.3cm]{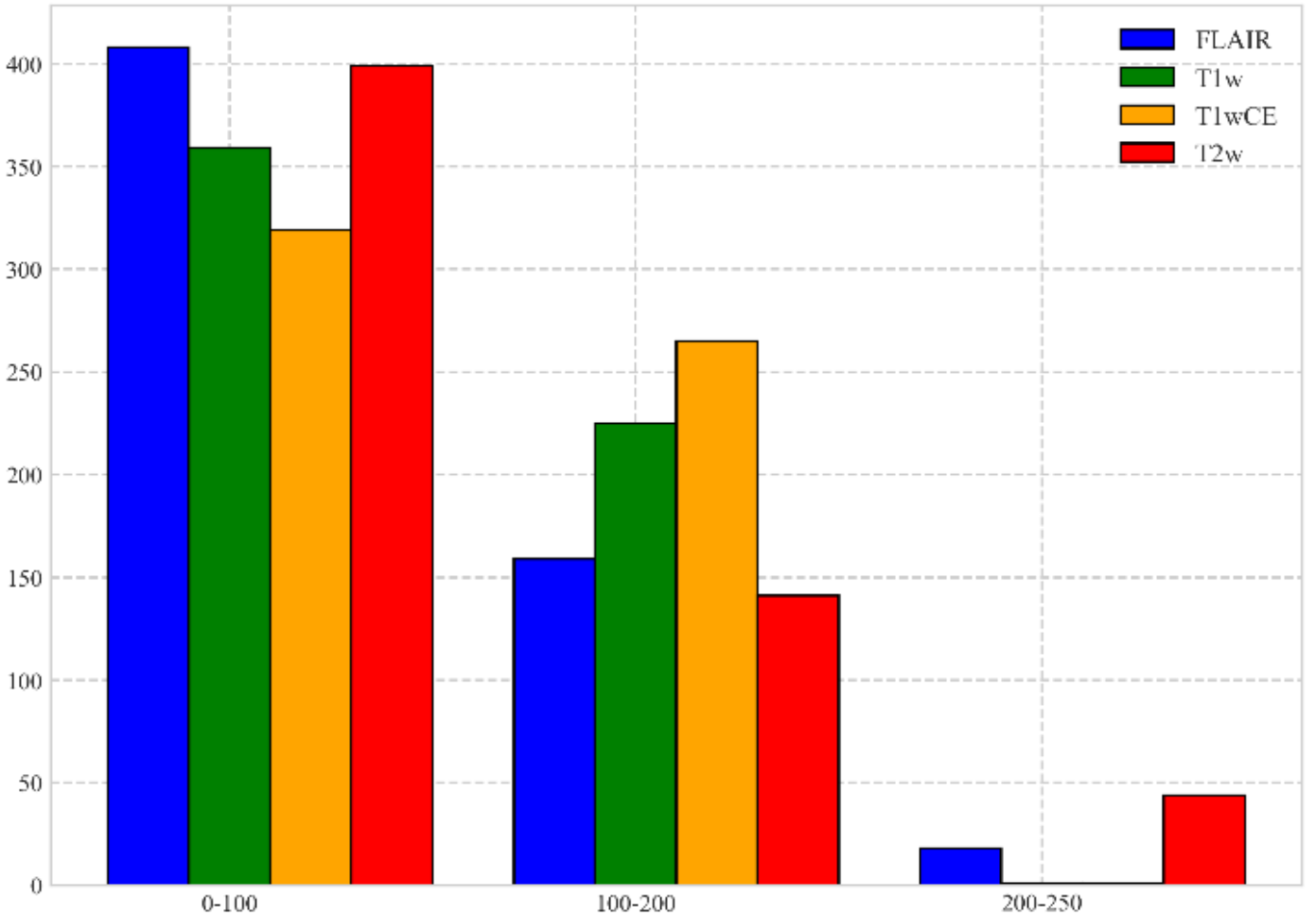}
    \caption{The total number of slices, for each modality, within each mpMRI scan.}
    \label{numbers}
\end{figure}

At first we performed segmentation of each slice of all mpMRI scans (for all modalities) so as to detect the brain regions and crop them. Then we removed some slices, mainly at the start and end of the MRI scan sequence of slices, since these slices displayed only a minuscule portion of the brain or were entirely void of any useful content. Fig. \ref{numbers} shows the total number of slices within each mpMRI scan (for each modality).
Finally, all segmented slices were  resized  to $224 \times 224 \times 3$ pixel resolution and their intensity values were normalized  to  $[-1,1]$. The resulting segmented slices constitute the input to our method.

\subsection{Performance Metrics}
The performance measure is the average F1 Score across the two categories (i.e., macro F1 Score):

\begin{equation} \label{expr}
\mathcal{P} = \frac{\sum_{expr} F_1^{expr}}{2}
\end{equation}

The $F_1$ score is a weighted average of the recall (i.e., the ability of the classifier to find all the positive samples) and precision (i.e., the ability of the classifier not to label as positive a sample that is negative). The $F_1$ score  takes values in the range $[0,1]$; high values are desired. The $F_1$ score is defined as:

\begin{equation} \label{f1}
F_1 = \frac{2 \times precision \times recall}{precision + recall}
\end{equation}

\subsection{Implementation Details}
We used ResNet18 (discarding the output layer but keeping the global average pooling) as CNN and a single one-directional LSTM consisting of 128 units as RNN. The dense layer in the Routing component consisted of 64 hidden units and the dense layer in the Modality Fusion component consisted of 128 hidden units. We used batch size equal to 4 and lengths $t1$ and $t4$ (corresponding to FLAIR and T2 modalities, respectively) equal to 250, lengths $t2$ and $t3$ (corresponding to T1w and T1wCE modalities, respectively) equal to 200. We used stratified 5-fold cross validation. Training was divided in two phases: at first each modality stream/subnetwork was trained independently and then the multimodal method was trained end-to-end. We utilized SGD with momentum equal to 0.9 and SAM; learning rate was $10^{-4}$ when training from scratch and $10^{-5}$ when training in an end-to-end manner. Training was performed on a Tesla V100 32GB GPU.

\section{Experimental Results}

This section describes a set of experiments evaluating the performance  of the proposed approach. 
At  first, we compare the performance of BTDNet with the performance of the state-of-the-art methods that participated in the RSNA-ASNR-MICCAI BraTS 2021 Challenge; we show that it outperforms these networks by large margins. 
Finally, we perform ablation studies that illustrate the contribution of the various components of BTDNet. In particular, we focus on the: 3D Analysis component (i.e., the choice of CNN and RNN models); Routing component (i.e., the existence of the Mask layer, as well as the number of hidden units in the dense layer); contribution of each modality; data augmentations (geometric transformations, MixAugment and Test-Time Data Augmentation); objective function.

\subsection{Comparison with the state-of-the-art}
At first we compare the performance of BTDNet to that of the state-of-the-art methods, the unimodal 3D-Resnet10-Trick (winner of RSNA-ASNR-MICCAI BraTS 2021 Challenge), the multimodal EfficientNet-LSTM-mpMRI (runner up of the Challenge), the multimodal EfficientNet-Aggr (third place in the Challenge), the unimodal YOLO-EfficientNet (fourth place in the Challenge), the multimodal stats-EfficientNet (fifth place in the Challenge). 
For fair comparison with our method which utilized segmented 3D volumes, we re-implemented these methods and also utilized the same segmeneted data.
Table \ref{sota} shows the performance of the state-of-the-art methods, as well as our method's performance. The results are denoted in the form of: average of 5 folds $\pm$ spread (i.e., difference between max and min performance within the 5 folds).
It can be seen from Table \ref{sota} that our method outperforms all other state-of-the-art methods by large margins; in more detail it outperforms the winner of the Challenge, 3D-Resnet10-Trick, by 3.3\% and it outperforms all other methods by at least 18.8\%. Table \ref{sota} also shows that our method achieves the minimum spread among all state-of-the-art methods.

Let us note that all state-of-the-art methods utilized the ad-hoc strategy of selecting fixed input length by removing or duplicating slices within each 3D volume. Let us further note that some methods utilized CNN-RNN architectures in which the 3D volume annotation was propagated to each slice within the volume. Finally, we performed an extra experiment which we do not show in Table \ref{sota} so as to not clutter the results. In more detail, we implemented 3D-Resnet10-Trick (winner of the Challenge) and utilized non-segmented data. Surprisingly we noticed that the method's performance increased from $62.9 \pm 4.8$ (that it was when segmented data were used) to $63.3 \pm 4.4$; this indicates that the method focused more on regions outside of the brain which played a role in the increase in the performance (however this is not correct as the MGMT promoter methylation status exists in the brain region and should be predicted only from that region).

\begin{table}[h]
\captionsetup{justification=centering}
\caption{Comparison between BTDNet and the state-of-the-art on the RSNA-ASNR-MICCAI BraTS 2021 Challenge dataset}
\centering
\scalebox{1.}{
\begin{tabular}{cc}
\hline
Methods                                 &  F1 Score
\\ \hline \hline
stats-EfficientNet	&	42.9 $\pm$ 6.84 \\
YOLO-EfficientNet  & 44.4 $\pm$ 7.41      \\ 
EfficientNet-Aggr  & 46.1 $\pm$ 4.68 \\
EfficientNet-LSTM-mpMRI & 47.4 $\pm$ 4.41 \\
3D-Resnet10-Trick  & 62.9 $\pm$ 4.8  \\ \hline
\textbf{BTDNet} & \textbf{66.2} $\pm$ \textbf{3.1}   \\ \hline
\end{tabular}
}
\label{sota}
\end{table}

\subsection{Ablation Study}
In the following, we perform various ablation experiments to evaluate the contribution of each novel component of BTDNet. 
At first, we utilize different backbone networks to act as the CNN in the 3D Analysis component of BTDNet. 
We experimented with using ResNet50 and ResNet101 (which are bigger networks than ResNet18 which is chosen as the CNN in BTDNet), EfficientNetB0 and EfficientNetB3 (as has been used by state-of-the-art methods and have proved their value), as well as ConvNeXt-T \cite{liu2022convnet}. 
As can be seen in Table \ref{ablation}, ResNet18 is a better backbone CNN than its bigger counterparts (ResNet50 and ResNet101); we can see that the bigger the ResNet networks gets (and the more layers it has), the lower the performance is. This occurs probably due to the small size of the training dataset; larger models are too complex to model it. This also occurs in the case of the EfficientNet models as the performance of EfficientNetB0 is higher than that of EfficientNetB3. Table \ref{ablation} also illustrates that the performance of ConvNeXt-T is higher than that of ResNet50, which was expected. Finally, overall best performance has been achieved when ResNet18 is utilized.
As a second ablation experiment, we utilize different  networks to act as the RNN in the 3D Analysis component of BTDNet. We experimented with using LSTM and GRU, also varying their number of units from 64 to 128 and 256. As can be seen in Table \ref{ablation}, LSTM and GRU produce similar performance. Best results have been obtained when LSTM with 128 units had been used. If the number of units were increased, the performance of the method started to decrease.

Next, we assess the contribution of the Routing component and its subcomponents. In more detail, we experimented with not using the Routing component at all or without using the Mask Layer and thus our method was tackling the problem with the different lengths of each volume as a traditional ad-hoc strategy by either removing slices when a larger length is met  or duplicating slices when the volume contains a smaller number of slices. Table \ref{ablation} proves that these approaches do not yield good results; our method outperforms them by large margins (at least 3.3\%), as was the case with the state-of-the-art methods that used such ad-hoc strategies. This was expected as removing of slices results in loss of important information  for the final decision and duplication of slices negatively
affects the final decision and makes the method gets biased towards the repeating data and thus towards one category. We further experimented with using different number of hidden units in the dense layer of the Routing component; we experiments with using 32, 64 and 128 units. Best results have been obtained when we used 64 units.

\begin{table}[h]
\captionsetup{justification=centering}
\caption{Ablation Study for the contribution of each component of BTDNet on the RSNA-ASNR-MICCAI BraTS 2021 Challenge dataset}
\centering
\scalebox{1.}{
\begin{tabular}{cc}
\hline
BTDNet                                 &  F1 Score
\\ \hline \hline
ResNet50 as CNN	&	64.1 $\pm$ 3.6 \\
ResNet101 as CNN	&	62.8 $\pm$ 3.88 \\
EfficientNetB0 as CNN	&	63.9 $\pm$ 3.92 \\
EfficientNetB3 as CNN	&	62.9 $\pm$ 4.21 \\
ConvNeXt-T as CNN &	64.9 $\pm$ 3.7 \\ 
\hline \hline
LSTM, 64 units  as RNN &	65.1 $\pm$ 3.5 \\
LSTM, 256 units  as RNN &	64.3 $\pm$ 4.3 \\
GRU, 64 units  as RNN &	65 $\pm$ 3.7 \\
GRU, 256 units  as RNN &	64.4 $\pm$ 4.4 \\ 
\hline \hline
no Routing component  &	60.9 $\pm$ 5.7 \\
no Mask Layer &	62.9 $\pm$ 5.2 \\
Dense layer (Routing component), 32 units &	65.1 $\pm$ 3.4 \\
Dense layer (Routing component), 128 units &	64.9 $\pm$ 3.6 \\
\hline \hline
only FLAIR modality &	64.5 $\pm$ 3.3 \\
only T1w modality &	63.2 $\pm$ 4.3 \\
only T1wCE modality &	63.8 $\pm$ 4.1 \\
only T2 modality &	64.1 $\pm$ 3.8 \\
\hline \hline
no Test-Time Data Augmentation  &	65.2 $\pm$ 3.4 \\
no geometric transformations &	65.3 $\pm$ 3.3 \\
no MixAugment &	64.7 $\pm$ 3.8 \\
\hline \hline
Categorical Cross Entropy as Objective Function &	64.9 $\pm$ 3.4 \\
Binary Cross Entropy as Objective Function &	64.5 $\pm$ 3.5 \\
\hline \hline
\textbf{BTDNet} & \textbf{66.2} $\pm$ \textbf{3.1}   \\ \hline
\end{tabular}
}
\label{ablation}
\end{table}

In the following, we assess the contribution of each modality and thus of the multimodal approach. Table \ref{ablation} illustrates the results that our method achieves when it is a unimodal approach (using either the FLAIR, or the T1w, or the T1wCE, or the T2 modalities) and when it is a multimodal approach (that utilizes all available modalities). One can see in Table \ref{ablation}, that among the unimodal methods, when the FLAIR modality is used, best performance is achieved; when T2 modality is used, second best performance is achieved; when T1wCE modality is achieved, it outperforms the case when T1w is used. Table \ref{ablation} shows that when the multimodal approach is used, it outperforms all unimodal approaches.
This was expected as different modalities provide complementary information. By incorporating multiple modalities, our method could access a richer and more diverse set of data, leading to a more comprehensive understanding of the targeted problem. Additionally, our multimodal method could probably reduce redundancy and ambiguity in data interpretation, becoming more robust and less susceptible to errors or noise. Finally, in some cases, there could be limited data available in a single modality and thus a multimodal approach can help mitigate data scarcity issues by leveraging information from other modalities.

In the following, we perform an ablation study in regards to the proposed data augmentation strategies. In more detail, Table \ref{ablation} shows the performance of our method when: i) no Test-Time Data Augmentation had been performed (i.e., during inference the method was evaluated only on the real non-augmented data and three augmented versions of the data had not been created and used in the evaluation); ii) no geometric transformations had been used (i.e. both random horizontal flip and rotation had not been performed on the data during training); iii) no MixAugment had been used (i.e., the method was trained only using the real training data and during training no virtual data had been constructed and subsequently used along with the real data). 
It can be seen that in all cases, the performance of our method decreases, verifying that each data augmentation strategy alone brings a performance gain, and best performance is achieved when all are used together. The biggest decrease in performance is observed when  the MixAugment strategy is not used. Let us also note that we performed one more experiment which we do not show to not clutter the results; we performed only the geometric transformations (i.e. random horizontal flip and rotation) and the same transformation has been performed on the slices withing the same volume. Our method's performance in this case was 0.8\% lower than its corresponding performance when a different random geometric transformation has been performed on the slices withing the same volume.

Finally, the last ablation experiment is with respect to the contribution of the objective function. We experiment with using the proposed multi-class focal loss and the standard categorical cross entropy (when there are two outputs in the method) and binary cross entropy (when there is one output in the method) losses. Table \ref{ablation} shows that when multi-class focal loss was utilized, our method achieved the best performance, which was at least 1.3\% higher than the corresponding one when the other losses were used.

%%%%%%%%%%%%%%%%%%%%%%%%%%%%%%%%%%%%%%%%%%
\section{Conclusions}

In this paper we proposed BTDNet, a new multimodal approach which harmonizes analysis of 3-D image volumes consisting of different number of slices and annotated per volume. In more detail, BTDNet accepts as input a mpMRI (i.e., a FLAIR 3D volume, a T1w 3D volume, a T1wCE 3D volume and a T2 3D volume) and predicts the MGMT promoter methylation status. BTDNet consists of three components: the 3D Analysis, the Routing, the Modality Fusion and the Data Augmentation ones. 

When BTDNet is fed with a new 3D volume input, at first a CNN performs local (per 2-D image/slice) analysis, extracting features from each slice. Then these features are fed to an RNN, placed on top of the CNN, so as to capture their temporal dependencies. The CNN and RNN networks perform global (per 3D volume) analysis. The RNN output features are then concatenated and fed to a Mask layer which dynamically selects RNN outputs taking into account their input length. The output of the Mask layer is then fed to a dense layer. This procedure is followed for each of the four modalities; the output features of the dense layer for each modality are then concatenated and fed to a dense layer that maps them to the same feature space. Finally, the output layer follows providing the final classification. The objective function for training BTDNet was based upon the Focal Loss for multi-class classification.

Excellent performance has been achieved on the dataset of the  RSNA-ASNR-MICCAI BraTS 2021 Challenge verifying our developments and surpassing all state-of-the-art methods by large margins. Finally, an extensive ablation study has been conducted and presented illustrating the novel contributions of this work.
%

%Future work includes extension of \textcolor{red} to include uncertainty estimation and domain adaptation to a large variety of other databases and related applications.  

\clearpage

\bibliographystyle{splncs04}
\bibliography{refs}
\end{document}